\documentclass[aps,twocolumn,noshowkeys]{revtex4-2}
\usepackage{dcolumn,amsmath}
\usepackage{graphicx}
\graphicspath{ {./Images/} }
\usepackage{bm} 
\usepackage{hyperref}     

\begin{document}

\title{Actinide and lanthanide molecules to search for strong CP-violation}

\author{L.V.\ Skripnikov$^{1,2,*}$, N.S.\ Mosyagin$^{1}$, A.V.\ Titov$^{1}$}
\affiliation{$^{1}$Petersburg Nuclear Physics Institute named by B.P.\ Konstantinov of National Research Center ``Kurchatov Institute'' (NRC ``Kurchatov Institute'' - PNPI),
1 Orlova roscha mcr., Gatchina, 188300 Leningrad region, Russia
}
\affiliation{$^{2}$Saint Petersburg State University, 7/9 Universitetskaya nab., St. Petersburg, 199034 Russia}
\homepage{http://www.qchem.pnpi.spb.ru    }

\email{\\ skripnikov\_lv@pnpi.nrcki.ru, leonidos239@gmail.com}

\author{V.V.\ Flambaum$^{3,4}$}
\affiliation{$^{3}$School of Physics, The University of New South Wales, Sydney NSW 2052, Australia}
\affiliation{$^{4}$Johannes Gutenberg-Universit\"at Mainz, 55099 Mainz, Germany}

\date{23.03.2020}

\begin{abstract}
 The existence of the fundamental CP-violating interactions inside the nucleus leads to the existence of the nuclear Schiff moment. 
The Schiff moment potential corresponds to the electric field localized inside the nucleus and directed along its spin. This field can interact with electrons of an atom and induce the permanent electric dipole moment (EDM) of the whole system. The Schiff moment and corresponding electric field are enhanced in the nuclei with the octupole deformation leading to the enhanced atomic EDM. There is also a few-order enhancement of the T,P-violating effects in molecules due to the existence of energetically close levels of opposite parity. We study the Schiff moment enhancement in the class of diatomic molecules with octupole-deformed lanthanide and actinide nuclei: $^{227}$AcF, $^{227}$AcN, $^{227}$AcO$^+$, $^{229}$ThO, $^{153}$EuO$^+$ and $^{153}$EuN. Projecting the existing experimental achievements to measure EDM in diamagnetic molecules with spherical nucleus ($^{205}$TlF) to the considered systems one can expect a very high sensitivity to the quantum chromodynamics  parameter ${\bar \theta}$ and other hadronic CP-violation parameters surpassing the current best limits by several orders of magnitude. It can have a dramatic impact on the modern understanding of the nature of CP-violating fundamental interactions.
\end{abstract}

\maketitle

\section{Introduction}

Search of the time-reversal (T) and spatial parity (P) violation  effects is one of the most important probes for theories beyond the Standard Model~\cite{Safronova:18}~\cite{Note1}.
In particular, it can shed light on the matter-antimatter asymmetry~\cite{sakharov1967violation} problem. Nonzero permanent electric dipole moments (EDM's) of elementary particles, atoms, and molecules imply manifestation of the existence of the T,P-violating interactions. The strongest limit on the electron EDM has been established in experiments on the beam of \textit{paramagnetic} $^{232}$ThO molecules~\cite{ACME:18}. This limit is almost two orders of magnitude stronger than that obtained in the best atomic-type experiment on paramagnetic Tl atoms~\cite{Regan:02}. Another type of experiment to search for the electron EDM has been performed using the trapped molecular cations $^{180}$HfF$^+$~\cite{Cornell:2017} which also allows to achieve sensitivity surpassing atomic~\cite{Regan:02} one.

Corresponding experiments with \textit{diamagnetic} atoms and molecules are most sensitive to the T,P-violating nuclear forces which can also induce EDM of the whole system.  The strongest limit on the \textit{atomic} EDM has been obtained for the diamagnetic $^{199}$Hg atom~\cite{Hg2016}. Experiments are also performed on $^{225}$Ra~\cite{Bishof:2016}, $^{129}$Xe~\cite{Allmendinger:2019} and Rn~\cite{Rn} atoms. Molecules are very promising systems for such experiments as they can be fully polarized in laboratory electric fields due to 
the
existence of close levels of opposite parity. However, the only such experiment with the \textit{diamagnetic molecule} has been performed on $^{205}$TlF~\cite{Cho:91}. A new ``cold molecule nuclear time reversal experiment'' (CeNTREX)  with this molecule is now under construction~\cite{CeNTREX,Norrgard:2017}. It  aims to measure a shift in the nuclear magnetic resonance frequency of the thallium nuclei when the molecules are polarized~\cite{CeNTREX}. The expected sensitivity of this experiment is about 3 orders of magnitude higher than that in the previous one~\cite{CeNTREX}.

The contribution of the electron EDM in the
diamagnetic systems is strongly suppressed. According to the Schiff theorem \cite{Schiff:63,Sushkov:84,Flambaum:86} the nuclear EDM is screened by electrons and cannot contribute to the EDM of a neutral system (see also book \cite{Khriplovich:91} and references therein).
Therefore, T,P-violating EDM's of neutral diamagnetic atoms and molecules are mostly induced by the interaction of the nuclear Schiff moment~\cite{Schiff:63,Sushkov:84} with electrons. Experiments on Hg and TlF utilize spherical nuclei. However, nuclei with the octupole deformation can have much larger Schiff moments~\cite{Auerbach:1996,Spevak:97}. The enhancement is due to the collective nature of the intrinsic moments and the small energy separation between members of parity doublets in such nuclei~\cite{Auerbach:1996,Spevak:97} (see Appendix). Schiff moment can be induced by different T,P-violating mechanisms inside the nucleus. Therefore, it is possible to express the T,P-violating atomic or molecular effect in terms of the fundamental parameters of the interactions such as the quantum chromodynamics (QCD) parameter ${\bar \theta}$ (which is connected to the strong CP problem) as well as other hadronic CP-violation parameters~\cite{Flambaum:19a,Flambaum:2020a,Engel:2003,Dobaczewski:2018}. 
An accurate electronic structure calculation of the atom (molecule) is required for this to connect corresponding atomic (molecular) effect with the nuclear Schiff moment by studying its interaction with electrons, or, by other words, to calculate the enhancement factor determined by the electronic structure. Note, that this factor cannot be measured.

 Nuclei with collective octupole deformation are available in a number of isotopes of Fr, Rn, Ra and light actinide atoms, also in some lantanide isotopes. These isotopes would have incomplete shells if the nucleus is spherical. This is why the  minimum of energy is achieved for a different shape. An idea of the octupole deformation may be explained by the fact that light nuclei have a bigger binding energy per nucleon than heavy nuclei. Therefore, some energy gain may be achieved if we make pear-shaped heavy nucleus from two overlapping  tightly bound nucleon clusters.
 
The largest Schiff moments were predicted for the nuclei of the lanthanide and actinide atoms ($f$--elements) ~\cite{Flambaum:19a,Flambaum:2020a}. However, all previous experiments, as well as \textit{ab-initio} studies of the Schiff moment enhancement in molecules have been performed only for molecules containing $s$-- and $p$--elements~\cite{Petrov:02,Quiney:98b,Hinds:80a,Parpia:97,Kudrin:2019,Kudashov:13,Kudashov:14,Skripnikov:16a}.
Some estimations of the Schiff moment enhancement in molecules were made in Ref.~\cite{Flambaum:19a} for molecules containing $f$--elements based on the atomic estimates. But as we show in the present paper, the uncertainty of such estimations can be rather large. This is due to the limited applicability of the picture in which heavy atom of the diatomic molecule is treated as an ion in a (uniform) external electric field. 
Besides, the Z-scaling of the molecular enhancement parameters between molecules containing elements belonging to different groups of  Mendeleev's Periodic Table is not guaranteed (Z is the nuclear charge).
 For an accurate treatment the consideration of the manifold of (nonlinear) chemical bonding formation effects is required. 
Qualitatively, one should consider here possible repulsion effects between electrons participating in the chemical bond formation and electrons of a lone pair, that lead to opposite sign contributions~\cite{Kudashov:13}, etc. All such effects should be treated only within the explicit molecular calculation which takes into account both correlation and relativistic effects at a high level of molecular theory. 

In the present paper we accurately study the Schiff moment enhancement for the class of molecules containing f--elements: $^{227}$AcF, $^{227}$AcN, $^{227}$AcO$^+$, $^{229}$ThO, $^{153}$EuO$^+$ and $^{153}$EuN. The expected T,P-violating effect for these systems is expressed in terms of the QCD parameter ${\bar \theta}$. For comparison, the $^{205}$TlF molecule has been also studied at the same level of theory. Note that the T,P--violating effect in molecules, which we considered in the current work, is more than 2 orders of magnitude larger than that in TlF. The structure of the contributions of various sources of T,P-violation in the systems under consideration is very different from that in the case of TlF.
It means that additional experiments with proposed molecules will allow one to set restrictions on different fundamental parameters 
more strictly,
i.e. without the suggestion that there is only one source of the symmetry violation. Finally, corresponding nuclei are stable or have very large lifetimes and are available in macroscopic quantities.  Therefore, it follows from our study that the experiments on the considered molecules can lead to significant improvements of the limits on hadronic CP-violation parameters or even result in the non-zero values. In both cases, this will have
a dramatic impact on the modern understanding of the nature of CP-violating fundamental interactions.

\section{Theory}

The 
nuclear Schiff moment $\mathbf{S}$ is defined by the following expression~\cite{Sushkov:84}:
\begin{equation}
{\bf S}=\frac{e}{10} [<r^2 {\bf r}> - \frac{5}{3Z}<r^2><{\bf r}>], 
  \label{SchiffMoment}
\end{equation}
where $e$ is the electron charge,
$<r^n> \equiv \int \rho_{nuc}({\bf r}) r^n d^3r$ are the moments of the nuclear charge density $\rho_{nuc}$ and ${\bf r}$ is measured from the nuclear center-of-mass position. Vector $\mathbf{S}$ is directed along the nuclear spin. 
Nuclei with the octupole deformation have large intrinsic collective Schiff moment, proportional to the collective octupole moment ~\cite{Auerbach:1996,Spevak:97}. All odd electric moments (including electric dipole, octupole and  Schiff moments) vanish in the laboratory frame if parity is conserved.  Indeed, EDM and Schiff moment are polar $T$-even vectors which must be directed along the nuclear spin $I$ which is $T$-odd pseudovector.  
This vanishing happens due the nuclear rotation which makes average orientation of the nuclear axis zero, $<{\bf k}>=0$, and all odd moments correlated with this axis do not show up in the laboratory frame. However, time and parity violating nuclear forces mix nuclear rotational states of opposite parity (which form a doublet for non-zero nuclear spin $I$, similar to the $\Lambda$-doubling in molecules) and produce orientation of the nuclear axis ${\bf k}$ along the nuclear spin, $<{\bf k}> \propto {\bf I}$. This makes the electric dipole and Schiff moments directed along the nuclear spin in the laboratory frame, $<{\bf S}> \propto {\bf I}$ \cite{Auerbach:1996,Spevak:97}.

For a spherical nucleus with one unpaired nucleon, both terms in Eq.~(\ref{SchiffMoment}) are comparable in absolute value but have opposite signs. Both of them often are not known accurately. This can lead to the large uncertainty of the Schiff moment for such a nucleus. The problem with the cancellation does not arise for the Schiff moment of a nucleus with octupole deformation since the second term in Eq.~(\ref{SchiffMoment}) is strongly suppressed.
Indeed, if the shape of the proton and neutron distributions is the same, the intrinsic electric dipole moment relative to the centre of mass vanishes, $e <{\bf r}>=0$.  Absence of the cancellation makes the result stable.  A detailed description of the Schiff moment calculation can be found in Refs.~\cite{Flambaum:19a,Flambaum:2020a} and references therein.
A few equations explaining the origin and magnitude of the Schiff moment ~\cite{Auerbach:1996,Spevak:97,Flambaum:2020a} are also given in the Appendix.

\begin{figure}[]
\centering
\includegraphics[width = 3.3 in]{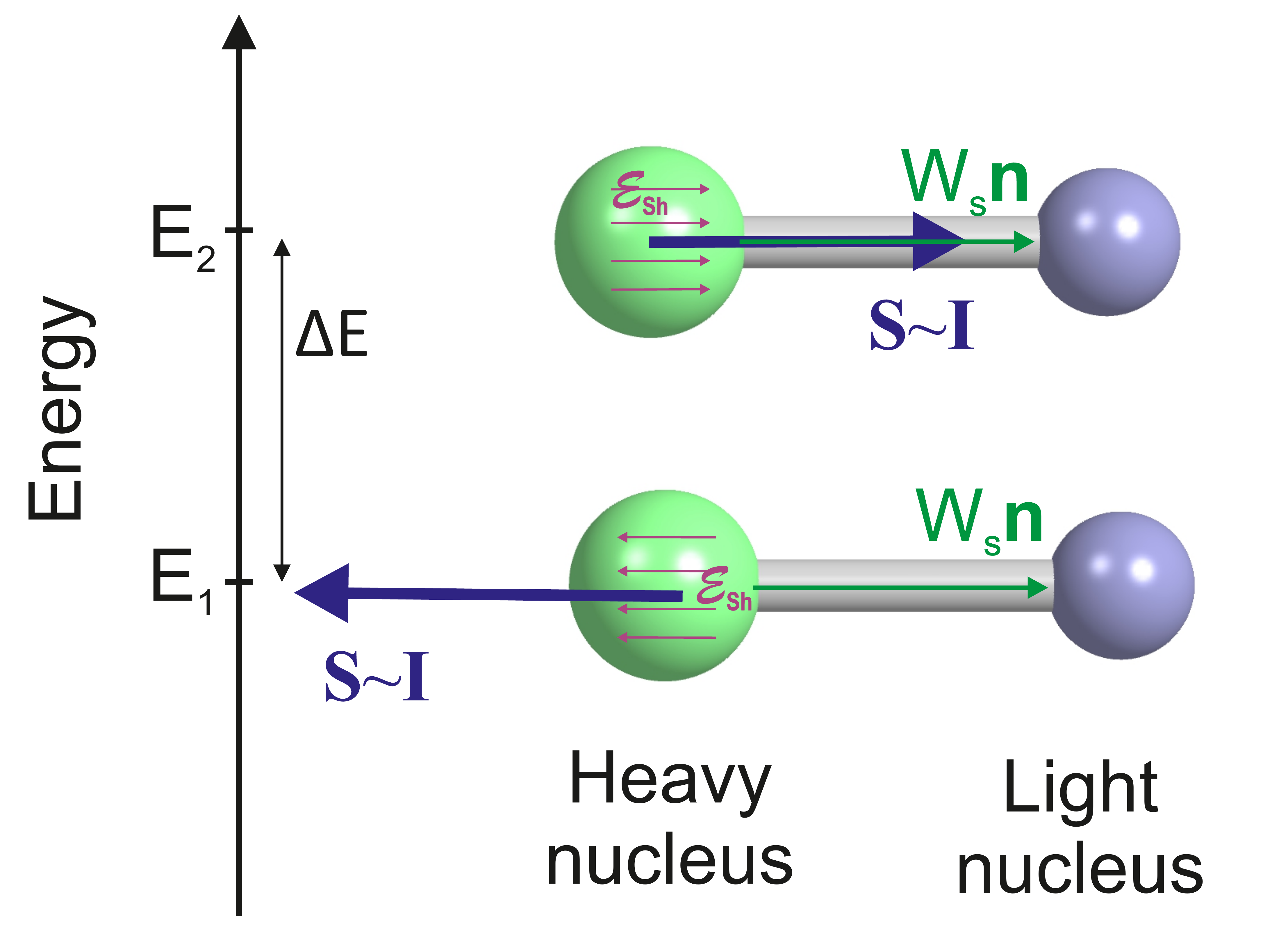}
 \caption{The nuclear Schiff moment in a diatomic molecule. There is the T,P-violating energy shift $\Delta E$ between two configurations with the mean value of the Schiff moment $\mathbf{S}$ directed parallel or anti-parallel to the molecular axis $\mathbf{n}$. Approximately constant electric field of the Schiff moment $\mathcal{E}_{Sh}$ is localized inside the nucleus and directed along the nuclear spin $\mathbf{I}$.}
 \label{FigMol}
\end{figure}

The T,P-violating effect caused by the Schiff moment in case of a diatomic molecule is described by the following effective Hamiltonian \cite{Hinds:80a,Sushkov:84}:
\begin{equation}
     H^{{\rm eff},0}= 6 X \mathbf{S} \cdot \mathbf{n}\ = W_S^{(0)} \mathbf{S} \cdot \mathbf{n},
 \label{HX}
\end{equation}
where $\mathbf{n}$ is the unit vector directed along the internuclear axis (axis $z$) from a heavy atom to a light one, $W_S^{(0)}=6 X$ and $X$ is determined by the electronic structure of the molecule under consideration:
\begin{eqnarray}
X=-\frac{2\pi}{3}\langle\Psi|[\sum_i \mathbf{\nabla_i}\cdot \mathbf{n},\delta(\mathbf{R})]|\Psi\rangle \\
=\frac{2\pi}{3} \mathbf{n} \cdot \mathbf{\nabla} \rho_{e}\left(\mathbf{r}\right)|_\mathbf{R}
  \label{X}
\end{eqnarray}
where the sum is over all electrons, $\Psi$ is the electron wave function, $\mathbf{R}$ is the heavy nucleus position,
$\rho_{e}(\mathbf{r})$ is the electronic density calculated from $\Psi$.
The effective Hamiltonian in Eqs.~(\ref{HX})--(\ref{X}) misses finite nuclear size corrections which are significant for heavy nuclei. 
The effective Hamiltonian for the case of a finite-size nucleus is~\cite{Ginges:04,KozlovA:2012}:
\begin{equation}
     H^{{\rm eff},2}= W_S^{(2)} \mathbf{S'} \cdot \mathbf{n},
 \label{HXnew}
\end{equation}
where $S'$ is the corrected nuclear Schiff moment~\cite{KozlovA:2012} and $W_S^{(2)}$ is defined in the following way:
\begin{equation}
    W_S^{(2)}=\langle\Psi|\sum_i  \frac{3\mathbf{r_i} \cdot \mathbf{n}}{B}\rho_{nuc}|\Psi\rangle,
  \label{W_S}
\end{equation}
where 
$B=\int\rho_{nuc}(r)r^4dr$.
Expressions~(\ref{HXnew}) and (\ref{W_S}) suggest the existence of approximately constant electric field, $\mathcal{E}_{Sh}$, which is localized inside the nucleus and directed along the nuclear spin~\cite{Ginges:04} (see Fig.~\ref{FigMol}). 
In 
the
case of
the
free atom, this nuclear field polarizes its electronic structure and produces atomic EDM.

\begin{figure}[]
\centering
\includegraphics[width = 3.3 in]{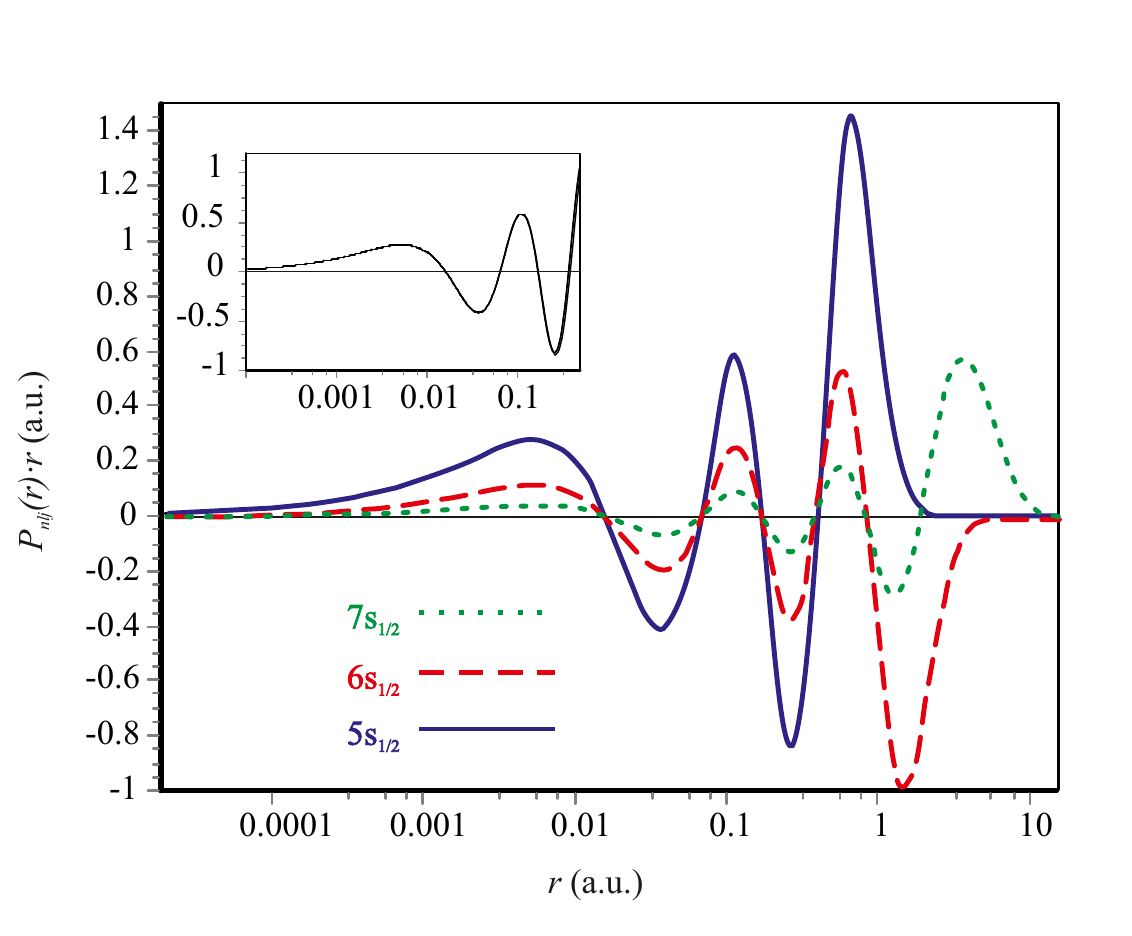}
 \caption{Radial parts of large components of the $5s_{1/2}$, $6s_{1/2}$ and $7s_{1/2}$  spinors of Th for the $7s^27p^16d^1$ configuration.  Inset: Large components of $5s_{1/2}$, $6s_{1/2}$ and $7s_{1/2}$ spinors in the core region; scaling factors are chosen in such a way that the amplitudes of large components of these spinors are equal at $R_c$ = 0.25~Bohr.}
 \label{PropFig}
\end{figure}

From the property of proportionality of the one-electron wavefunctions in the vicinity (or inside) the heavy atom nucleus (see Fig.\ref{PropFig}) corresponding matrix elements of operators whose action is concentrated in this region are proportional to each other~\cite{Khr91,Titov:99,Dzuba:2011,Skripnikov:15b,Flambaum:2019b}.
Therefore, one has $W_S^{(2)}= W_S^{(0)}/r^{sp}$. The proportionality coefficients $r^{sp}$ can be calculated analytically~\cite{Flambaum:2019b}.
In the present paper, the $X$ parameters have been calculated in accordance with previous molecular studies. Parameters $W_S^{(2)}$ are obtained by applying the $r^{sp}$ factors.

It was shown in Refs.~\cite{Titov:06amin, Petrov:02,Skripnikov:11a,Skripnikov:15b} that the relativistic four-component problem of evaluating matrix elements such as Eq.(\ref{X}) can be effectively divided into two steps. For this, the space around a given heavy atom is divided into valence and core regions. In
the first step, one calculates the molecular wave function using the 
generalized 
relativistic effective core potential (GRECP) Hamiltonian \cite{Titov:99, Mosyagin:10a,Mosyagin:16,Mosyagin:17}. It is built in such a way that the corresponding wave function is very accurate in the valence region but has incorrect behavior in the core region. At the second step, the true four-component behavior of the wave function is restored in the core region using the procedure~\cite{Titov:06amin, Petrov:02,Skripnikov:11a,Skripnikov:15b,Titov:96b,Skripnikov:14a,Skripnikov:16a} based on a proportionality of valence and virtual (unoccupied in the reference Slater determinant) spinors in the inner-core region of the heavy atom (see Fig.\ref{PropFig}).
Note that at the restoration step wavefunctions are represented by power series of the electronic radius-vector inside the nucleus. This allows one to eliminate complications in reproducing the asymptotic wavefunction behavior in the region near the nucleus~\cite{Skripnikov:16a}. The latter is especially important to calculate matrix element (\ref{X}) for which one has a strong cancellation of the large and small component contributions~\cite{Quiney:98b}.

The many-body problems of calculating wave functions $\Psi$ for the molecules under consideration have been solved using the ``all-order'' method with respect to single and double excitations, in which some of the most important connected triple excitations are also taken into account, i.e., the coupled cluster with single, double, and perturbative treatment of triple cluster amplitudes, CCSD(T)~\cite{Stanton:97}.
The 30-electron, 29-electron, 21-electron, and 35-electron valence GRECPs \cite{Mosyagin:16,Mosyagin:17} were used for 
an
accurate description of the valence and outer-core electrons of Th, Ac, Tl, and Eu atoms, respectively. In correlation calculations, all these electrons as well as all electrons of the light atoms were included.
We have constructed the uncontracted basis set for the Ac atom containing 20\,$s$--, 20\,$p$--, 10\,$d$--, 8\,$f$--, 5\,$g$--, 3\,$h$-- and 2\,$i-$type Gaussian basis functions, which can be written as Ac[20,20,10,8,5,3,2]. Basis functions of  $g$--, $h$-- and $i$-- types have been obtained using the method of constructing natural basis sets~\cite{Skripnikov:13a}. Basis sets Th[20,15,15,10,6,5,2], Tl[23,16,10,9,3], Eu[14,14,10,8,3,2] were constructed in a similar way.  Uncontracted Dyall's AETZ basis sets from Ref.~\cite{Dyall:2016} were used for the O, F and N atoms.
 
For molecular calculations we used codes from Refs.~\cite{DIRAC15,CFOUR}. The code developed in Refs. \cite{Petrov:02,Skripnikov:11a,Skripnikov:15b} has been employed to calculate $X$ parameters.

\section{Results and discussion}

Table~\ref{TResdip} gives the values of the equilibrium internuclear distances, that were used in calculations of the $X$ constants. Experimental values for the distances are available only for ThO and TlF molecules~\cite{Huber:79,Edvinsson:84}. For other molecules and cations, equilibrium distances have been obtained theoretically within the scalar-relativistic CCSD method. Table~\ref{TResdip} also gives the values of the molecule-frame dipole moments calculated at the two-component CCSD(T) level. A good agreement is found between the theoretical and available experimental values (as well as previous theoretical value for the dipole moment of the ground state of ThO~\cite{Buchachenko:10}).

\begin{table}[!h]
\caption{Equilibrium internuclear distances ($R_e$) and the absolute value of the molecule-frame dipole moment ($\mu$) with respect to the center of mass. Where available, the experimental values are given in brackets}
\begin{tabular}{llll}
\hline
Mol.     & State                &  $R_e$, Bohr {   } & $\mu$, Debye  \\
\hline
AcF      & $^1\Sigma^+$           &  4.00         &  2.2       \\
AcN      & $^1\Sigma^+$           &  3.61         &  7.6       \\
AcO$^+$  & $^1\Sigma^+$           &  3.56         &  7.0       \\
ThO      & $^1\Sigma^+$           &  3.47 (3.478~\cite{Huber:79,Edvinsson:84})  &  2.8  (2.782(12) \cite{Steimle:2011})   \\
EuO$^+$  & $(f^6)$ $^{*}$       &  3.32$^{*}$    &  5.8$^{*}$ \\
EuN      & $(f^6)$ $^{*}$       &  3.28$^{*}$    &  7.8$^{*}$ \\
TlF      & $^1\Sigma^+$          &  3.94  (3.93893(39)~\cite{Boeckh1964})&  4.1 (4.2283(8)~\cite{Boeckh1964})   \\
\hline
\end{tabular}

$^{*}$ 
The
spin-orbit part of the GRECP operator has been omitted in the calculation. Therefore, we give only 
the
configuration of the molecular state.

\label{TResdip}
\end{table}

Calculated values of the $X$ and  $W_S^{(2)}$ constants for the molecules and cations under consideration are given in Table~\ref{TRes1}. The most accurate (final) results were obtained within the two-component (i.e. including the spin-orbit interaction) CCSD(T) method. Table~\ref{TRes1} also presents values calculated at the Hartree-Fock (HF) and CCSD levels for comparison. One can see that correlation effects  strongly contribute to $X$ (e.g. about 27\% in the case of AcF). The largest contribution (5\%) of perturbation triple cluster amplitudes is for the AcN molecule.

\begin{table*}
\caption{Molecular constants $X$ and $W_S^{(2)}=6 X/r^{sp}$ ($e/a_B^4$, $a_B$=1 Bohr) calculated at different levels of theory, given in square brackets}
\begin{tabular*}{\textwidth}{@{\extracolsep{\fill}}llrrrrr}
\hline
Mol.     & State                &  X           & X           & X           & $r^{sp}$  & $W_S^{(2)}$   \\
         &                      & [HF]         & [CCSD]      & [CCSD(T)]   &           & [CCSD(T)]     \\         
\hline
AcF      & $^1\Sigma^+$           &  -2022       & -1569       & -1593       & 1.16      & -8240         \\
AcN      & $^1\Sigma^+$           &  -10580      & -9415       & -8950       & 1.16      & -46295        \\
AcO$^+$  & $^1\Sigma^+$           &  -13362      & -11600      & -11302      & 1.16      & -58461        \\
ThO      & $^1\Sigma^+$           &   -3965      &  -3187      & -3332       & 1.17      & -17085        \\
EuO$^+$  & $(f^6)$ $^{*}$       &  -2475$^{*}$ & -2140$^{*}$ & -2114$^{*}$ & 1.09      & -11677$^{*}$  \\
EuN      & $(f^6)$ $^{*}$       &  -1975$^{*}$ & -1847$^{*}$ & -1890$^{*}$ & 1.09      & -10419$^{*}$  \\
 TlF      & $^1\Sigma^+$          &  9111        &  7262       &  7004       & 1.13      & { }37192      \\
\hline
\end{tabular*}
$^{*}$ 
The
spin-orbit part of the GRECP operator has been omitted in the calculation. Therefore, we give only 
the
configuration of the molecular state.
\label{TRes1}
\end{table*}

The final value of $X$(TlF) is in a good agreement with the previous correlation calculation performed in Ref.~\cite{Petrov:02} while the Hartree-Fock value is also in good agreement with Refs. \cite{Quiney:98b,Hinds:80a,Parpia:97}.
Theoretical uncertainty of the $X$ values for the AcF, AcN, AcO$^+$, ThO, and TlF molecules was estimated
with a procedure similar to that given in Ref.~\cite{Kudashov:13} and is about 10\%. 
 The total electronic angular momentum of the ground electronic state of the Eu$^{3+}$ cation is zero. Therefore, one can expect that
 the
 corresponding electronic state of EuN and EuO$^+$ molecules with $f^6$ configuration and zero projection of the total electronic angular momentum on the molecular axis  will be the ground or metastable one. Due to difficulties of describing the molecular configuration with six open-shell $f$-electrons it was not possible to theoretically determine the ground electronic states of EuN and EuO$^+$ and the spin-orbit part of the GRECP operator has been turned off in these calculations. 
 However, the $f$-type electrons have negligible amplitudes inside the nucleus and practically do not contribute to matrix element (\ref{X}) which is of the main interest here. Thus, the detailed description of $f$-type electrons 
  is not very
 important for 
  $X$ calculation and one can estimate the uncertainty of $X$(EuO$^+$) and $X$(EuN) as 15\% taking also into account missed spin-orbit contribution~\cite{Skripnikov:16a}.

As it can be seen from Table~\ref{TRes1}, $X$(AcF)=-0.2$X$(TlF) is rather far from the estimation of $X$(AcF)=3.5$X$(TlF) expected from simple atomic-based rescaling~\cite{Flambaum:19a}. This suggests that the explicit molecular electronic structure calculation is required to obtain reliable values for molecular constants. One can also see that $X$(AcN)=5.6$X$(AcF). Qualitatively, this can be explained as follows. One electron in AcF goes (is polarized) from Ac towards the F atom resulting in the Ac$^{+}$ cation. However, Ac$^{+}$ has also two other valence electrons which can go in the opposite direction from the former electron. This leads to a partial cancellation of the contributions to $X$. One can see from Table~\ref{TResdip} that molecule-frame dipole moments of AcF and AcN show the same trend.

The difference between the values of $X$(AcO$^+$) and $X$(AcN) is smaller than between $X$(AcN) and $X$(AcF), but not negligible: $X$(AcO$^+$)=1.2$X$(AcN). Note that the difference in constants of T,P-violating interactions a neutral molecule, and isoelectronic cation can be even larger~\cite{Skripnikov:15b}.

In the experiment it is necessary to work with completely polarised molecules to achieve all the benefits from the molecular enhancement~\cite{CeNTREX,Norrgard:2017}. A special technique has been developed to work with molecular cations~\cite{Cornell:2017}. For neutral molecules in the $^1\Sigma^+$ state the characteristic electric field that is required to polarise a molecule is of order $2B_e/\mu$, where $B_e=\frac{\hbar^2}{2MR_e^2}$ is the rotational constant and M is the reduced mass. It follows from Table~\ref{TResdip} that the characteristic polarising field for the cases of ThO, AcF should be about twice larger than in 
the
case of TlF. For AcN and EuN molecules the field should be approximately the same as in the case of TlF molecule.
Knowledge of the polarising field is important for the experiment planning and preparation.

Schiff moment is induced by the CP-violating forces inside the nucleus. The dominating contribution to these forces is due to the $\pi$-meson exchange (the $\eta$-meson exchange can also contribute~\cite{Flambaum:19a,Flambaum:2020a}). Therefore, the Schiff moment can be expressed in terms of the 
strong $\pi$-meson–nucleon interaction constant $g$ and 
$\pi$-meson–nucleon CP-violating interaction constants 
${\bar g}_0$, ${\bar g}_1$, ${\bar g}_2$.
These constants can be further expressed in terms of more fundamental quantities such as the QCD parameter ${\bar \theta}$ or the quark chromo-EDMs ${\tilde d_u}$ and ${\tilde d_d}$~\cite{Bsaisou:15,Vries:2015,Yamanaka:2017}
(see also Appendix).
Thus, it is possible to obtain $S({\bar \theta})$. The compilation of the latest nuclear estimations as well as new estimates of $S({\bar \theta})$ are given in~\cite{Flambaum:19a,Flambaum:2020a}. The strongest  limit on ${\bar \theta}$ follows from the neutron EDM and Hg EDM experiments~\cite{Hg2016,Engel:2013}: ${\bar \theta} < 10^{-10}$. One can use this limit and dependence of $S({\bar \theta})$ as well as calculated in the present paper $W_S$ molecular constants to estimate the expected effect for the molecules and ions under consideration. For a fully polarised molecule, the energy difference which can be measured in experiments is (see Fig.~\ref{FigMol}):
\begin{equation}
     \Delta E=2W_S S.
 \label{DeltaE}
\end{equation}
Table~\ref{TRes2} gives the $\Delta E$ values for the molecules under consideration and ${\bar \theta} = 10^{-10}$.

\begin{table}
\caption{Estimated energy shift $\Delta E=2 W_S S({\bar \theta})$ for ${\bar \theta}=10^{-10}$. $S({\bar \theta})$ dependencies of the corresponding heavy nuclei are taken from Ref.~\cite{Flambaum:19a}. $W_S$ values are taken from Table~\ref{TRes1} above}
\begin{tabular*}{0.48\textwidth}{@{\extracolsep{\fill}}llrr}
\hline
Molecule & State &  $S$,                   & {                } $|\Delta E|$,     \\
         &       &  $e$ fm$^3 {\bar \theta}$  & mHz     \\
\hline
$^{227}$AcF      & $^1\Sigma^+$              & 6     & 0.4    \\
$^{227}$AcN      & $^1\Sigma^+$              & 6     & 2.5    \\
$^{227}$AcO$^+$  & $^1\Sigma^+$              & 6     & 3.1    \\
$^{229}$ThO       & $^1\Sigma^+$              & $\leq 2$ & $\leq 0.3$ \\
$^{153}$EuO$^+$  & $(f^6)$             & $-3.7$ & 0.4   \\
$^{153}$EuN      & $(f^6)$               & $-3.7$ & 0.3   \\
$^{205}$TlF      & $^1\Sigma^+$                & $0.02$ & 0.007   \\
\hline
\end{tabular*}
\label{TRes2}
\end{table}

The experiment to measure the Schiff moment of the $^{205}$Tl nucleus has been undertaken for the $^{205}$TlF molecule in 1991~\cite{Cho:91}. The measured energy shift was~$(-0.14\pm 0.24)$ mHz~1991~\cite{Cho:91}. The new CeNTREX experiment with this molecule is now under construction and is expected to achieve three orders of magnitude higher sensitivity already in its first generation~\cite{CeNTREX,Norrgard:2017}. 
Therefore, taking into account estimations in Table~\ref{TRes2} one can conclude that the use of the similar experimental technique for the considered molecules seems to be very promising to set new limits (or measure) the QCD ${\bar \theta}$ parameter and other hadronic CP-violation parameters
~\cite{Note2}.

It is necessary to have several experiments to unambiguously separate different contributions of the T,P-violating effects to the observed molecular effect (see, e.g.~\cite{Jung:13,Skripnikov:17c}).
Structure of such contributions 
are very different in case of TlF with the spherical nucleus and in the proposed systems with octupole-deformed nuclei. For Tl one has the following contribution to the Schiff moment from ${\bar \theta}$,  ${\tilde  d}_d$ and ${\tilde  d}_u$~\cite{Flambaum:86,Flambaum:19a,Flambaum:2020a}:
\begin{align}\label{Tlschiff}
 \nonumber
 S(^{203}{\rm Tl},{\bar \theta})  \approx  S(^{205}{\rm Tl},{\bar \theta})  \approx 0.02  \,{\bar \theta} \, e\cdot\textrm{fm}^3 \ ,\\
 S(^{203}{\rm Tl},{\tilde d})  \approx  S(^{205}{\rm Tl},{\tilde d}) \approx (12 {\tilde d}_d +9 {\tilde d}_u)\, e\cdot \textrm{fm}^2 \, .
 \end{align}
The most accurate calculations for the deformed nucleus $^{225}$Ra give~\cite{Engel:2003,Flambaum:19a,Flambaum:2020a}:
\begin{align}
 S(^{225}{\rm Ra}, {\bar \theta})& \approx  - \,{\bar \theta} \, e\cdot \textrm{fm}^3 \ , \label{Raschiff} \\
 S(^{225}{\rm Ra},{\tilde d})& \approx  10^4 ( 0.50 \,{\tilde d}_u -  0.54 \,{\tilde  d}_d )\, e\cdot \textrm{fm}^2 \, . \nonumber
 \end{align} 
Note that in Refs.~\cite{Flambaum:19a,Flambaum:2020a} similar ratio of contributions from different sources was implied for other octupole-deformed nuclei. Therefore, experiments on the proposed molecules with the deformed nuclei are complementary to the current experiment on TlF.

Note, that the $^{153}$Eu nucleus is stable, while $^{227}$Ac and $^{229}$Th have very large lifetimes: 21.8 and 7900 years, respectively. All of the considered nuclei are available in macroscopic quantities. From this point of view experiments with considered molecules can be performed easier than with the $^{225}$RaO~\cite{Kudashov:13}: though $^{225}$Ra has also enhanced Schiff moment~\cite{Engel:2003,Auerbach:1996,Spevak:97} its lifetime is 14.9 days.


\section*{Acknowledgements}
We are grateful to Laura McKemmish and Mikhail Kozlov for useful discussions.
Electronic structure calculations in the paper have been carried out using computing resources of the federal collective usage centre Complex for Simulation and Data Processing for Mega-science Facilities at NRC ``Kurchatov Institute", http://ckp.nrcki.ru/, and computers of Quantum Chemistry Lab at NRC ``Kurchatov Institute" - PNPI
L.V.S. and V.V.F. express a special thanks to the New Zealand Institute for Advanced Study for its hospitality and support of scientific discussions. N.S.M. is grateful to the governor of Leningrad district for the personal scientific fellowship on the GRECP generation. \\

Molecular electronic structure calculations have been supported solely by the Russian Science Foundation Grant No. 19-72-10019. The calculations of the nuclear structure were supported by the Australian Research Council Grant No. DP150101405 and New Zealand Institute for Advanced Study.


\appendix
\section{Appendix: Schiff moment estimation}
It may be instructive to present few equations explaining the origin and magnitude of the Schiff moment ~\cite{Auerbach:1996,Spevak:97,Flambaum:2020a}.
If a nucleus has an octupole deformation $\beta_3$ and a quadrupole deformation $\beta_2$, in the fixed-body  (rotating) frame the Schiff moment $S_{intr}$ is proportional to the octupole moment $O_{intr}$, i.e. it  has a collective nature:
\begin{equation}\label{Sintr}
 S_{intr} \approx \frac{3}{5 \sqrt{35}} O_{intr} \beta_2 \approx    \frac{3}{20 \pi  \sqrt{35}} e Z R^3 \beta_2 \beta_3 ,
\end{equation}
where $R$ is the nuclear radius.
Deformation parameters for different nuclei are compiled e.g. in Ref.~\cite{Moller2016}.
Nucleus with an octupole deformation and non-zero nucleon angular momentum  has a doublet of close opposite parity rotational states $|I^{\pm}>$ with the same angular momentum $I$ ($| I^{\pm} >=\frac{1}{\sqrt{2}} (|\Omega> \pm |-\Omega>)$, where $\Omega$ is  the projection of $I$ onto the nuclear axis). The states of this doublet are mixed by $P,T-$violating interaction $W$. The mixing coefficient is:
\begin{equation}\label{alpha}
 \alpha_{+-}=\frac{<I^-| W| I^+>}{E_+  -  E_-}, 
\end{equation}
where $E_+$ and  $E_-$ are energies of the opposite parity rotational states in the $\Omega$- doublet.
This mixing polarises  nuclear axis ${\bf k}$ along the nuclear spin ${\bf I}$, $<k_z>= 2 \alpha _{+-}\frac{I_z}{I+1}$,
and the intrinsic Schiff moment shows up in the laboratory frame:
\begin{equation}\label{Scol}
 S= 2 \alpha_{+-} \frac{I}{I+1} S_{intr}. 
\end{equation}
   According to Ref. ~\cite{Spevak:97} the T,P-violating matrix element is approximately equal to
   \begin{equation}\label{W}
   <I^-| W| I^+> \approx \frac{\beta_3 \eta}{A^{1/3}} [ \textrm{eV}].
  \end{equation}  
  Here
A is the number of the nucleons in the nucleus (atomic mass),
  $\eta$ is the dimensionless strength constant of the nuclear $T,P$- violating potential $W$:
   \begin{equation}\label{eta}
 W= \frac{G}{\sqrt{2}} \frac{\eta}{2m} ({\bf \sigma \nabla}) \rho ,
   \end{equation}
where $G$ is the Fermi constant, $m$ is the  nucleon mass, $\rho$ is the nuclear number density, $\sigma$ is the Pauli matrix and  $\nabla$ is the gradient operator.  Eqs. 
(\ref{Sintr})-(\ref{W}) 
give an analytical estimate for the  Schiff moment: 
\begin{equation}\label{San}
 S \approx 1. \cdot 10^{-4} \frac{I}{I+1} \beta_2 (\beta_3)^2 Z A^{2/3} \frac{[\textrm{KeV]}}{E_-  -  E_+} e \,\eta \, [\textrm{fm}^3],
   \end{equation}
This estimate is in agreement with more accurate numerical calculations available for  a number of nuclei ~\cite{Spevak:97}.

Within the meson exchange theory, the $\pi$-meson exchange gives the dominating contribution to the T,P-violating nuclear forces. In the standard notations $g$ is the strong $\pi$-meson - nucleon  interaction constant and ${\bar g}_0$, ${\bar g}_1$, ${\bar g}_2$ are the  $\pi$-meson - nucleon CP-violating interaction  constants in the isotopic channels $T=0,1,2$. One can express the results in terms of more fundamental parameters such as the QCD $\theta$-term constant  ${\bar \theta}$ and  the  quark chromo-EDMs ${\tilde d_u}$ and  ${\tilde d_d}$. In Ref. ~\cite{Flambaum:2020a} we presented the results of the substitutions in the following form:
\begin{align}\label{Sg} 
  S(g) & \approx  K_S(  - 2.6 g {\bar g}_0 +  12.9 g {\bar g}_1 -6.9 g {\bar g}_2)\, e\cdot \textrm{fm}^3\, ,\\
  \label{Stheta} 
  S({\bar \theta})& \approx  - K_S\,{\bar \theta} \, e\cdot \textrm{fm}^3 \ ,\\
\label{Sd} 
 S({\tilde d})& \approx  10^4 K_S ( 0.50 \,{\tilde d}_u -  0.54 \,{\tilde  d}_d )\, e\cdot \textrm{fm}^2 \, ,
\end{align}   
where $K_S=K_I K_{\beta}K_A K_E$, $K_I=\frac{3 I}{I+1}$, $K_{\beta}=791\beta_2 (\beta_3)^2$, $K_A=0.00031 Z A^{2/3}$,  $K_E= \frac{55\textrm{KeV}}{E^-  -  E^+} $. 
Numerical factors are chosen such that  these coefficients are equal to 1 for  $^{225}$Ra (where sophisticated many-body calculations ~\cite{Engel:2003} giving Eq. (\ref{Sg}) with $K_S=1$ were performed) and are of the order  of unity for other heavy nuclei with octupole deformation. All relevant nuclear parameters and the values of $K_S$ for deformed nuclei with strongly enhanced  collective Schiff moments  are presented in Ref. ~\cite{Flambaum:2020a}.
The accuracy of the analytical Schiff moment calculations is hardly better than a factor of 2. Future numerical many-body calculations similar to that for $^{225}$Ra should lead to the improvement  of the accuracy.

\end{document}